\documentclass[12pt]{article}
\usepackage{amsmath}
\usepackage{amssymb}
\usepackage{dsfont}
\usepackage[ruled]{algorithm2e}
\usepackage{fullpage}
\usepackage{pstricks}
\usepackage{amsthm}
\def\abs#1{{|\,#1\,|}}
\def\mirror{{\operatorname{Mir}}}
\def\id{{\operatorname{Idt}}}
\def\tr{{\operatorname{Trn}}}
\def\mtt#1{{\tt #1}}
\newtheorem{theorem}{Theorem}
\newtheorem{proposition}[theorem]{Proposition}

\newtheorem{lemma}[theorem]{Lemma}

\title{Pseudo-Power Avoidance}
\author{Ehsan Chiniforooshan \and Lila Kari \and Zhi Xu}
\date{The University of Western Ontario, \\
Department of Computer Science, \\
Middlesex College, London, Ontario, Canada, N6A 5B7 \\
{\tt \{ehsan,lila,zhi\underline{ }xu\}@csd.uwo.ca} \\
\medskip \today}
\begin{document}
\maketitle

\begin{abstract}
Repetition avoidance has been studied since Thue's work. In this
paper, we considered another type of repetition, which is called
pseudo-power. This concept is inspired by Watson-Crick
complementarity in DNA sequence and is defined over an antimorphic
involution $\phi$. We first classify the alphabet $\Sigma$ and the
antimorphic involution $\phi$, under which there exists sufficiently
long pseudo-$k$th-power-free words. Then we present algorithms to
test whether a finite word $w$ is pseudo-$k$th-power-free.
\end{abstract}

\section{Introduction}\label{section:introduction}
Let $\Sigma$ be an alphabet. The set of finite words and infinite
words over $\Sigma$ are denoted by $\Sigma^*$ and $\Sigma^\omega$,
respectively. Word $v$ is called a \emph{factor} of $x$ if $x=uvw$
for some words $u$ and $w$. A nonempty word $w$ is called a
\emph{square} if $w$ can be written as $w=uu$ for some
$u\in\Sigma^*$, and is called a \emph{cube} if $w$ can be written as
$w=uuu$ for some $u\in\Sigma^*$. For example, the English word
``murmur'' is a square. More generally, for an integer $k\geq 2$, a
nonempty word $w$ is called a \emph{$k$th power} if $w=u^k$ for some
$u\in\Sigma^*$. A word $w$ is called \emph{square-free}
(respectively, \emph{cube-free}, \emph{$k$th-power-free}) if $w$
does not contain any square (respectively, cube, $k$th power) as a
factor. In the early 1900's, Thue showed in a series of papers
examples of square-free infinite words over $3$ letters and $4$
letters respectively and cube-free infinite word over binary
alphabet \cite{Thue1906,Thue1912} (see \cite{Berstel1995} for
English translation of Thue's work). In 1921, Morse \cite{Morse1921}
independently discovered Thue's construction. In 1957, Leech
\cite{Leech1957} showed another construction of square-free infinite
word, which is generated by morphism. In 1995, Yu \cite{Yu1995}
showed square-free infinite words that cannot be generated by any
morphism.

Let $\abs{w}_a$ be the total number of letter $a$ that appear in the
word $w$, and let $\abs{w}=\sum_{a\in\Sigma}\abs{w}_a$ for
$w\in\Sigma^*$. A nonempty word $w$ is called an \emph{abelian
square} if $w$ is in the form $w=u_1u_2$ such that
$\abs{u_1}_a=\abs{u_2}_a$ for each letter $a$. For example, the
English word ``teammate'' is a abelian square. Analogously,
$w\Sigma^*$ is called an \emph{abelian cube} if $w=u_1u_2u_3$, where
$\abs{u_1}_a=\abs{u_2}_a=\abs{u_3}_a$ for each $a\in\Sigma$, and
called an \emph{abelian $k$th power} if $w=u_1u_2\cdots u_k$, where
integer $k\geq 2$, $\abs{u_i}_a=\abs{u_j}_a$ for $a\in\Sigma$ and
$1\leq i,j\leq k$. A word $w$ is called \emph{abelian-square-free}
(respectively, \emph{abelian-cube-free},
\emph{abelian-$k$th-power-free}) if $w$ contains no abelian square
(respectively, abelian cube, abelian $k$th power) as a factor. In
1957, Erd\"os \cite{Erdos1957} asked whether there exists an
abelian-square-free infinite word. The construction of such words
were given by Pleasants \cite{Pleasants1970} in 1970 over $5$
letters and by Ker\"anen \cite{Keranen1992} in 1992 over $4$
letters. Most recently, Ker\"anen \cite{Keranen2009} presented a
great many new abelian-square-free infinite words. In 1979, Dekking
\cite{Dekking1979} discussed abelian-$k$th-power-free infinite words
for $k\geq 3$.

The discussion on $k$th power is related to biology. Repeats of
certain segments in human DNA sequences may predict certain disease
\cite{Mirkin2007}. A variation on the $k$th power, called pseudo
$k$th power, is defined by antimorphic involutions, which are
generalizations of the famous Watson-Crick complementarity function
in biology. Other concepts in combinatorics on words have been
generalized in the setting of antimorphic involutions, such as
pseudo-primitivity \cite{Czeizler&Kari&Seki2009} and
pseudo-palindrome \cite{Kari&Mahalingam2009}. A variation on the
concept of the pseudo $k$th power has also appeared in tiling
problems (see \cite{Beauquier&Nivat1991,Braquelaire&Vialard1999}),
where the function involved is a morphism other than an
antimorphism.

In this paper, we will discuss the word that does not contain any
pseudo $k$th power as a factor. Here the pseudo power is defined by
antimorphic involutions. In Section~\ref{section:freeword}, we will
introduce the concept of pseudo-power-free words and discuss the
existence of such words over different setting of alphabet and
antimorphic involutions. In Section~\ref{section:decision}, we will
discuss the algorithms for the decision problem of
pseudo-power-freeness. At the end, we will summarize the results and
present open problems.

\section{Pseudo-power-free infinite words}\label{section:freeword}
Without loss of generality, in the following discussion we always
assume the letters are $\mtt0,\mtt1,\mtt2,\ldots$. The empty word is
denoted by $\epsilon$. A function $\theta:\Sigma^*\to\Sigma^*$ is
called an \emph{involution} if $\theta(\theta(w))=w$ for all
$w\in\Sigma^*$, and called an \emph{antimorphism} (respectively,
\emph{morphism}) if $\theta(uv)=\theta(v)\theta(u)$ (respectively,
$\theta(uv)=\theta(u)\theta(v)$). We call $\theta$ an
\emph{antimorphic involution} if $\theta$ is both an involution and
an antimorphism. For example, the classic Watson-Crick
complementarity in biology is an antimorphic involution over four
letters. The \emph{mirror image} is the function of reversing a word
as defined by $\mirror(a_1a_2\cdots a_n)=a_n\cdots a_2a_1$. Then the
mirror image over any alphabet is also an antimorphic involution. A
\emph{transposition} $(a,b)$ is a morphism that is defined by
$(a,b)(a)=b,(a,b)(b)=a,(a,b)(c)=c\textrm{ for }c\neq a,c\neq
b,a<b,a,b,c\in\Sigma$. One can verify that the mirror image and any
transposition commute, and two transpositions $(a_1,a_2)$ and
$(b_1,b_2)$ commute for $a_i\neq b_j,i,j\in\{1,2\}$. By definitions,
every antimorphic involution on the alphabet is a permutation of the
letters. Furthermore, we have the following proposition.

\begin{proposition}\label{prop:decomposition}
Let $\theta$ be an antimorphic involution over the alphabet
$\Sigma$. Then $\theta$ can be uniquely written as the composition
of transpositions with a mirror image
\begin{equation}\label{equationdecomp}
  \theta=(a_0,a_1)\cdot(a_2,a_3)\cdots(a_{2m-2},a_{2m-1})\cdot\mirror
\end{equation}
up to changing the order of composition, where $a_i\neq a_j$ for
$i\neq j$ and %b$m$ is the number of transpositions and
$m\geq 0$.
\end{proposition}
\begin{proof}
First we show that $\theta(a)$ is a single letter for any letter
$a$. By definitions, $\theta(w)=\theta(\epsilon
w)=\theta(\epsilon)\theta(w)$, so $\theta(\epsilon)=\epsilon$. For
any $a\in\Sigma$, since $\theta(\theta(a))=a$, we have
$\abs{\theta(a)}>0$ and $\abs{\theta(a)}<2$. Hence
$\abs{\theta(a)}=1$ for any $a\in\Sigma$.

Now we prove the existence of the decomposition in
Eq.~(\ref{equationdecomp}) for $\theta$ by induction on the size of
the alphabet. If $\abs{\Sigma}=1$, then $\theta(\mtt0)=\mtt0$ and
$\theta(\mtt0^m)=\mtt0^m=\mirror(\mtt0^m)$. So $\theta=\mirror$ and
Eq.~(\ref{equationdecomp}) holds. If $\abs{\Sigma}=2$, then either
$\theta(\mtt0)=\mtt0$ or $\theta(\mtt0)=\mtt1$. One can verify that
$\theta=\mirror$ when $\theta(\mtt0)=\mtt0$ and
$\theta=(\mtt0,\mtt1)\cdot\mirror$ when $\theta(\mtt0)=\mtt1$.
Suppose Eq.~(\ref{equationdecomp}) holds for any $\abs{\Sigma}<n$.
For $\abs{\Sigma}=n\geq 3$, either $\theta(\mtt0)=\mtt0$ or
$\theta(\mtt0)=a$ for some $a\neq\mtt0,a\in\Sigma$. If
$\theta(\mtt0)=\mtt0$, then by induction hypothesis the restriction
of $\theta$ on alphabet $\Sigma\setminus\{\mtt0\}$ can be written as
$(a_0,a_1)\cdots(a_{2m-2},a_{2m-1})\cdot\mirror$. So
$\theta=(a_0,a_1)\cdots(a_{2m-2},a_{2m-1})\cdot\mirror$ in this
case. If $\theta(\mtt0)=a$, then $\theta(a)=\mtt0$. By induction
hypothesis, the restriction of $\theta$ on alphabet
$\Sigma\setminus\{\mtt0,a\}$ can be written as
$(a_0,a_1)\cdots(a_{2m-2},a_{2m-1})\cdot\mirror$. So
$\theta=(\mtt0,a)\cdot(a_0,a_1)\cdots(a_{2m-2},a_{2m-1})\cdot\mirror$.
Therefore, Eq.~(\ref{equationdecomp}) holds.

To show the uniqueness of the form in Eq.~(\ref{equationdecomp}), we
first notice that every transposition is the inverse of itself.
Assume there is another decomposition
$\theta=(b_0,b_1)\cdot(b_2,b_3)\cdots(b_{2n-2},b_{2n-1})\cdot\mirror$,
where $b_i\neq b_j$ for $i\neq j$ and $n\geq 0$. Then
$\theta(b_0)=b_1$ and thus there is some $(a_p,a_{p+1})=(b_0,b_1)$.
Since $a_i\neq a_j$ for $i\neq j$, the order of composition in
Eq.~(\ref{equationdecomp}) can be arbitrarily changed. Hence we have
\begin{align*}
  & (b_2,b_3)\cdots(b_{2n-2},b_{2n-1})\cdot\mirror \\
  =&(b_0,b_1)\cdot\theta \\
  =&(a_0,a_1)\cdots(a_{p-2},a_{p-1})\cdot(a_{p+2},a_{p+3})\cdots(a_{2m-2},a_{2m-1})\cdot\mirror.
\end{align*}
Continuing this procedure, it follows that
$\mirror=(a'_0,a'_1)\cdots(a'_{2m'-2},a'_{2m'-1})\cdot\mirror$. By
the construction of $\{a_i\}$, however, we know $a'_i\neq a'_j$ for
$i\neq j$. So $m'=0$, since otherwise $\mirror(a'_0)=a'_1\neq a'_0$.
Therefore, $m=n$ and
$\{(b_0,b_1),\ldots,(b_{2n-2},b_{2n-1})\}=\{(a_0,a_1),\ldots,(a_{2m-2},a_{2m-1})\}$.
\end{proof}

For any antimorphic involution $\theta$ over $\Sigma$, we define
  $$\id(\theta)={\{a\in\Sigma: \theta(a)=a\}},
  \quad\tr(\theta)={\{a\in\Sigma: \theta(a)>a\}}.$$
The following proposition follows directly from
Proposition~\ref{prop:decomposition}.

\begin{proposition}\label{prop:id+2tr=absigma}
Let $\theta$ be an antimorphic involution over the alphabet
$\Sigma$. Then $\theta$ can be written as the composition of
$\abs{\tr(\theta)}$ distinct transpositions with a mirror image, and
\begin{equation}\label{equationid2tr}
  \abs{\id(\theta)}+2\abs{\tr(\theta)}=\abs{\Sigma}.
\end{equation}
\end{proposition}
\begin{proof}
By the proof of Proposition~\ref{prop:decomposition}, $\theta$ can
be written as
$\theta=(a_0,a_1)\cdot(a_2,a_3)\cdots(a_{2m-2},a_{2m-1})\cdot\mirror$,
where $a_i\neq a_j$ for $i\neq j$. Then
$\id(\theta)=\Sigma\setminus\{a_0,a_1,\ldots,a_{2m-1}\}$ and
$\tr(\theta)=\{a_0,a_2,\ldots,a_{2m-2}\}$. So $m=\abs{\tr(\theta)}$
and Eq.~(\ref{equationid2tr}) holds.
\end{proof}

For integer $k$ and antimorphism $\theta$, we call word $w$ a
\emph{pseudo $k$th power} (with respect to $\theta$) if $w$ can be
written as $w=u_1u_2\cdots u_k$, where either $u_i=u_j$ or
$u_i=\theta(u_j)$ for $1\leq i,j\leq k$. In particular, we call
pseudo $2$nd power by \emph{pseudo square}, and pseudo $3$rd power
by \emph{pseudo cube}. For example, over the alphabet $\{\mtt A,
\mtt T, \mtt C, \mtt G\}$ (here we use the conventional symbols
instead of $\{\mtt0,\mtt1,\mtt2,\mtt3\}$) with respect to the
Watson-Crick complementarity ($\mtt A\mapsto\mtt T, \mtt
T\mapsto\mtt A, \mtt C\mapsto\mtt G, \mtt G\mapsto\mtt C$), $\mtt
A\mtt C\mtt G\mtt C\mtt G\mtt T$ is a pseudo square and $\mtt A\mtt
C\mtt G\mtt T\mtt A\mtt C$ is a pseudo cube. By definitions, every
pseudo $k$th power is an abelian $k$th power with respect to the
mirror image, and every $k$th power is a pseudo $k$th power (with
respect to any antimorphic involution on the same alphabet). A word
$w$ is called \emph{pseudo-$k$th-power-free} (respectively,
\emph{pseudo-square-free}, \emph{pseudo-cube-free}) if $w$ cannot be
written as $w=uvx$ where $v$ is a pseudo $k$th power (respectively,
pseudo square, pseudo cube).

In the remaining of this section, we will discuss the following
problem: is there a pseudo-$k$th-power-free infinite word over
$\Sigma$ with respect to $\theta$? The discussion on
pseudo-power-free words is related to power-free words and
abelian-power-free words in the sense of the following lemmas.

\begin{lemma}\label{lemma:lemma0}
If $l$ is the minimal size of alphabet over which there is a
$k$th-power-free infinite word, then
\begin{itemize}
  \item[(1)] there is no pseudo-$k$th-power-free infinite word over $l-1$ or less letters;
  and
  \item[(2)] there is a pseudo-$k$th-power-free infinite word over $l'$ letters
  with respect to $\theta$, where $\abs{\tr(\theta)}\geq l$.
\end{itemize}
\end{lemma}
\begin{proof}
(1) If $l$ is the minimal size of alphabet over which there is a
$k$th-power-free infinite word, then for any $l''<l$ there is an
integer $N$ such that any word of length greater than $N$ over $l''$
letters contains a $k$th-power. Since a $k$th-power is a
pseudo-$k$th-power (with respect to any antimorphic involution), any
word of length greater than $N$ over $l''$ letters contains a
pseudo-$k$th-power with respect to $\theta$.

(2) Let $\theta$ be an antimorphic involution on $l'$ letters such
that $\abs{\tr(\theta)}\geq l$. We choose
$\Sigma'\subseteq\tr(\theta)$ such that $\abs{\Sigma'}=l$. Then
there is an infinite word $w$ over $\Sigma'$ such that $w$ is
$k$th-power-free. Now we claim that $w$ is also
pseudo-$k$th-power-free over $l'$ letters with respect to $\theta$.
Suppose $w$ contains a pseudo-$k$th-power. Then $w=xu_1u_2\cdots
u_ky$, where either $u_i=u_j$ or $u_i=\theta(u_j)$ for $1\leq
i,j\leq k$. For any $a\in\Sigma'$, by definition, $\theta(a)>a$ and
$\theta(\theta(a))=a<\theta(a)$. So $\theta(a)\not\in\Sigma'$ and
thus $\theta(u_i)$ is not a word over $\Sigma'$ for all $1\leq i\leq
k$. Hence $u_i=u_j$ for $1\leq i,j\leq k$ and $u_1u_2\cdots u_k$ is
a normal $k$th-power, which contradicts the fact that $w$ is
$k$th-power-free. Therefore, $w$ is pseudo-$k$th-power-free with
respect to $\theta$.
\end{proof}

\begin{lemma}\label{lemma:lemma1}
If there is an abelian-$k$th-power-free infinite word over $l$
letters, then there is a pseudo-$k$th-power-free infinite word over
$2l-1$ letters with respect to arbitrary antimorphic involution
$\theta$.
\end{lemma}
\begin{proof}
By Eq.~(\ref{equationid2tr}) in
Proposition~\ref{prop:id+2tr=absigma}, it follows that
$\abs{\tr(\theta)}<l$ and thus
$\abs{\id(\theta)}+\abs{\tr(\theta)}\geq l$. We choose
$\Sigma'\subseteq\id(\theta)\cup\tr(\theta)$ such that
$\abs{\Sigma'}=l$. Then there is an infinite word $w$ over $\Sigma'$
such that $w$ is abelian-$k$th-power-free. Now we claim that $w$ is
also pseudo-$k$th-power-free over $2l-1$ letters with respect to
$\theta$.

Suppose $w=xu_1u_2\cdots u_ky$ contains a pseudo-$k$th-power, where
either $u_i=u_1$ or $u_i=\theta(u_1)$ for $1\leq i\leq k$. Then
either $u_1$ is a word over $\id(\theta)$ or $u_1$ contains at least
one letter from $\tr(\theta)$. If $u_1\in\id(\theta)^*$, then
$\theta(u_1)=\mirror(u_1)$. So $u_1u_2\cdots u_k$ is an abelian
$k$th power, which contradicts the fact that $w$ is
abelian-$k$th-power-free. Otherwise, $u_1$ contains at least one
letter from $\tr(\theta)$, say $a$. Then since $\theta(a)>a$ and
$\theta(\theta(a))=a<\theta(a)$, we have
$\theta(a)\not\in\Sigma'\subseteq\id(\theta)\cup\tr(\theta)$. So
$u_i=u_1$ for $1\leq i\leq k$, and thus $w$ contains a $k$th power,
which again contradicts the fact that $w$ is
abelian-$k$th-power-free.

Therefore, $w$ is pseudo-$k$th-power-free with respect to $\theta$.
\end{proof}

\begin{lemma}\label{lemma:lemma2}
If there is a pseudo-$k$th-power-free infinite word over $\Sigma$
with respect to $\theta$, then there is a pseudo-$k$th-power-free
infinite word over $\Sigma'$ with respect to $\theta'$, where
$\abs{\tr(\theta')}\geq\abs{\tr(\theta)}$ and
$\abs{\id(\theta')}+\abs{\tr(\theta')}\geq\abs{\id(\theta)}+\abs{\tr(\theta)}$.
\end{lemma}
\begin{proof}
We choose $\Sigma_1\subseteq\tr(\theta')$ such that
$\abs{\Sigma_1}=\abs{\tr(\theta)}$. Since
$\abs{\id(\theta')}+\abs{\tr(\theta')}-\abs{\tr(\theta)}\geq\abs{\id(\theta)}$,
we can choose
$\Sigma_2\subseteq\id(\theta')\cup\tr(\theta')\setminus\Sigma_1$
such that $\abs{\Sigma_2}=\abs{\id(\theta)}$. Define
$\Sigma''=\Sigma_1\cup\{\theta'(a): a\in\Sigma_1\}\cup\Sigma_2$, and
define antimorphic involution $\theta''$ by $\theta''(a)=a,
\theta''(b)=\theta'(b)$ for $a\in\Sigma_2,
b\in\Sigma_1\cup\{\theta'(a): a\in\Sigma_1\}$. Then
$\abs{\Sigma''}=\abs{\Sigma}$,
$\id(\theta'')=\abs{\Sigma_2}=\abs{\id(\theta)}$ and
$\tr(\theta'')=\abs{\Sigma_1}=\abs{\tr(\theta)}$. So $\theta$ and
$\theta''$ are identical up to renaming of the letters. There is a
word $w$ over $\Sigma''$ such that $w$ is pseudo-$k$th-power-free
with respect to $\theta''$. We claim $w$ is also
pseudo-$k$th-power-free over $\Sigma'$ with respect to $\theta'$.

Suppose $w=xu_1u_2\cdots u_ky$ contains a pseudo-$k$th-power, where
either $u_i=u_1$ or $u_i=\theta'(u_1)$ for $1\leq i\leq k$. Then
either $u_1$ is a word over $\Sigma_1\cup\{\theta'(a):
a\in\Sigma_1\}\cup(\Sigma_2\cap\id(\theta'))$ or $u_1$ contains at
least one letter from $\Sigma_2\cap(\tr(\theta')\setminus\Sigma_1)$.
In the former case, $\theta''(u_1)=\theta'(u_1)$ and thus $w$
contains a pseudo-$k$th-power with respect to $\theta''$, which
contradicts to the pseudo-$k$th-power-freeness of $w$. In the latter
case, we assume $u_1$ contains
$a\in\Sigma_2\cap(\tr(\theta')\setminus\Sigma_1)$. One can verify
that $\theta'(a)\not\in\Sigma''$, so $u_i\neq\theta'(u_1)$ for all
$1\leq i\leq k$. Hence $w$ contains a $k$th-power $u_1u_2\cdots
u_k$, which again contradicts the pseudo-$k$th-power-freeness of
$w$.

Therefore, $w$ is pseudo-$k$th-power-free over $\Sigma'$ with
respect to $\theta'$.
\end{proof}

\subsection{Pseudo-square-free infinite words}
First we consider pseudo-square-free infinite words. Since every
binary word of length greater than $3$ contains squares, there is no
square-free infinite word over $2$ letters. By Ker\"anen's
construction of abelian-square-free infinite words, there exist
pseudo-square-free infinite words over $4$ letters with respect to
the mirror image. Furthermore, we have the following result.

\begin{proposition}\label{prop:pseudoleech}
For $3$-letter-alphabet, pseudo-square-free infinite word exists
with respect to mirror image and does not exist with any other
antimorphic involution.
\end{proposition}
\begin{proof}
There are two kinds of antimorphic involutions over $3$ letters:
$\theta$ is either the mirror image or a transposition composed with
the mirror image.

Suppose $\theta$ is the mirror image. Then the following morphism
$l$ given by Leech \cite{Leech1957} preserves square-freeness and
presents an infinite word $l^\omega(\mtt0)$, which is
pseudo-square-free with respect to the mirror image:
\begin{align*}
l(\mtt0)=&\mtt0\mtt1\mtt2\mtt1\mtt0\mtt2\mtt1\mtt2\mtt0\mtt1\mtt2\mtt1\mtt0,\\
l(\mtt1)=&\mtt1\mtt2\mtt0\mtt2\mtt1\mtt0\mtt2\mtt0\mtt1\mtt2\mtt0\mtt2\mtt1,\\
l(\mtt2)=&\mtt2\mtt0\mtt1\mtt0\mtt2\mtt1\mtt0\mtt1\mtt2\mtt0\mtt1\mtt0\mtt2.
\end{align*}
To see $w=l^\omega(\mtt0)$ contains no pseudo square, first we
observe that $w$ contains no square. If $w$ contains pseudo square
of the form $u_1u_2$ such that $u_1=\theta(u_2)$, then
$u_1u_2=u_2^Ru_2$ contains a square of length $2$ in the middle.
Since $w$ is square free, $w$ does not contain pseudo square.

Now suppose $\theta$ is a transposition composed with the mirror
image. Without loss of generality, we assume
$\theta=(\mtt0,\mtt1)\cdot\mirror$. We prove no pseudo-square-free
infinite word exists in this setting. Suppose $w$ is a
pseudo-square-free infinite word. A pseudo-square-free word (with
respect to $\theta$) cannot contains $\mtt0\mtt0, \mtt1\mtt1,
\mtt2\mtt2, \mtt0\mtt1, \mtt1\mtt0, \mtt2\mtt0\mtt1\mtt2,
\mtt2\mtt1\mtt0\mtt2$. So $w$ is either in
$(\mtt0\mtt2+\mtt1\mtt2)^\omega$ or in
$(\mtt2\mtt0+\mtt2\mtt1)^\omega$. If we omit all symbol $\mtt2$ in
$w$, the new infinite word should also be pseudo-square-free. But
the new word is over $\{\mtt0,\mtt1\}$ and every binary infinite
word contains pseudo squares, which is a contradiction. So there is
no pseudo-square-free infinite word over $3$ letters (with respect
to a transposition composed with the mirror image).
\end{proof}

There is another way to show the non-existence of pseudo-square-free
infinite word over $3$ letters with respect to $(\mtt0,\mtt1)$. We
use computer to find the longest pseudo-square-free word, if any.
Starting from empty word $\epsilon$, if a word is
pseudo-square-free, then we extend the word by adding a new letter
$\mtt0$ at the end; otherwise, we do back-tracking and try the next
letter. In other words, we do a depth-first-search in a labeled
tree, where each node presents a finite word. Such tree is called
\emph{trie} and application of similar technique has been appeared
in the literatures (for example, see \cite{Shallit2004}). In the
case of $3$ letters and $\theta=(\mtt0,\mtt1)\cdot\mirror$, the tree
is finite and all pseudo-square-free words are enumerated. There are
in total $91$ nodes, including $61$ leaves. The tree is of depth $8$
and one of the longest pseudo-square-free words is
$\mtt0\mtt2\mtt1\mtt2\mtt0\mtt2\mtt1$.

\begin{theorem}
There is no pseudo-square-free infinite word over $k$ letters for
$k\leq 2$, and there is a pseudo-square-free infinite word over $k$
letters for $k\geq 5$, with respect to arbitrary antimorphic
involution. For $3\leq k\leq 4$, the existence of pseudo-square-free
infinite words depends on the antimorphic involution.
\end{theorem}
\begin{proof}
(1) Since $3$ is the minimal size of alphabet over which there is a
square-free infinite word, by Lemma~\ref{lemma:lemma0}, there is no
pseudo-square-free infinite word over $k$ letters for $k\leq 2$ and
there is pseudo-square-free infinite word with respect to $\theta$
for $\tr(\theta)\geq 3$.

(2) Since there exists an abelian-square-free infinite word over $4$
letters, by Lemma~\ref{lemma:lemma1}, there is a pseudo-square-free
infinite word over $7$ or more letters.

(3) By Proposition~\ref{prop:pseudoleech}, over $3$ letters, there
is a pseudo-square-free infinite word with respect to mirror image,
where $\abs{\id(\mirror)}=3,\abs{\tr(\mirror)}=0$, and there is no
pseudo-square-free infinite word with respect to other antimorphic
involution. So by Lemma~\ref{lemma:lemma2}, there is a
pseudo-square-free infinite word with respect to $\theta$ such that
$\abs{\id(\theta)}+\abs{\tr(\theta)}\geq 3$.

The result is summarized in Table~\ref{table:pseudosquare}, where
the subscription presents which situation of the (1), (2), (3) the
case falls in.
\end{proof}

\begin{table}
\centering \caption{The existence of pseudo-square-free infinite
words}
\begin{tabular}{c|p{5.5ex}p{5.5ex}p{5.5ex}p{5.5ex}p{5.5ex}p{5.5ex}p{5.5ex}p{5.5ex}}
  \hline
  $\abs{\Sigma}$ & $1$ & $2$ & $3$ & $4$ & $5$ & $6$ & $7$ & $8$ \\
  \hline
  $\tr(\theta)=0$ & $\times_{1}$ & $\times_{1}$ & $\surd_{3}$ & $\surd_{3}$ & $\surd_{3}$ & $\surd_{3}$ & $\surd_{2,3}$ & $\surd_{2,3}$ \\
  $\tr(\theta)=1$ & $-$ & $\times_{1}$ & $\times_{3}$ & $\surd_{3}$ & $\surd_{3}$ & $\surd_{3}$ & $\surd_{2,3}$ & $\surd_{2,3}$ \\
  $\tr(\theta)=2$ & $-$ & $-$ & $-$ & {\small open} & $\surd_{3}$ & $\surd_{3}$ & $\surd_{2,3}$ & $\surd_{2,3}$ \\
  $\tr(\theta)=3$ & $-$ & $-$ & $-$ & $-$ & $-$ & $\surd_{1,3}$ & $\surd_{1,2,3}$ & $\surd_{1,2,3}$ \\
  $\tr(\theta)=4$ & $-$ & $-$ & $-$ & $-$ & $-$ & $-$ & $-$ & $\surd_{1,2,3}$ \\
  \hline
\end{tabular}\label{table:pseudosquare}
\end{table}

It is still a open problem that whether there exists a
pseudo-square-free infinite word over $4$ letters with respect to
antimorphic involution $\phi$, where $\tr(\phi)=2$. Experimental
computation shows that there are long pseudo-square-free words in
the setting, but we don't have proof of the existence of arbitrarily
long pseudo-square-free words. But if there are, they must satisfy
certain conditions as in the following proposition.

\begin{proposition}
$w$ is an pseudo-square-free infinite word over $4$ letters with
respect to $\phi$, where
$\tr(\phi)=2,\phi(a)=b,\phi(c)=d,a,b,c,d\in\{{\tt1,2,3,4}\}$, if and
only if $w\in ((a+b)(c+d))^\omega+((c+d)(a+b))^\omega$ and $w$ is
square free.
\end{proposition}
\begin{proof}
``$\Rightarrow$''.
% Suppose $w$ is an pseudo-square-free infinite word with respect to the antimorphic involution $\phi$.
Since $w$ is pseudo-square-free, $w$ is square free. Furthermore,
any word in $(a+b)^2+(c+d)^2$ is a pseudo square, so the letters in
$w$ must appear alternatively from $\{a,b\}$ and $\{c,d\}$. Hence
$w\in((a+b)(c+d))^\omega+((c+d)(a+b))^\omega$.

``$\Leftarrow$''.
% Suppose $w\in ((a+b)(c+d))^\omega+((c+d)(a+b))^\omega$ and $w$ is square free.
Suppose $w$ contains a pseudo square. Since $w$ is square free, it
must be the case that $w=ux\phi(x)u$. By the definition of
antimorphism involution, the last letter of $x$ and the first letter
of $\phi(x)$ are either both from $\{a,b\}$ or both from $\{c,d\}$,
which contradicts the fact $w\in
((a+b)(c+d))^\omega+((c+d)(a+b))^\omega$.
\end{proof}

\subsection{Other pseudo-power-free infinite words}
Now we consider pseudo-cube-free infinite words. By Dekking's
construction of abelian-cube-free infinite words, there exist
pseudo-cube-free infinite words over $3$ letters with respect to the
mirror image. The case over unary alphabet is trivial. For binary
alphabet, we have the following result.

\begin{proposition}\label{prop:pseudocube}
Pseudo-cube-free infinite word does not exist over binary alphabet
with any antimorphic involution.
\end{proposition}
\begin{proof}
There are two kinds of antimorphic involutions over binary alphabet:
we have either $\theta=\mirror$ or
$\theta=(\mtt0,\mtt1)\cdot\mirror$.

Suppose $\theta=\mirror$. Again, we use computer to find the longest
pseudo-cube-free word, if any. Starting from empty word $\epsilon$,
if a word is pseudo-cube-free, then we extend the word by appending
$\mtt0$; otherwise, we do back-tracking and try the next letter. The
resulted depth-first-search tree is finite. There are in total $171$
nodes, including $86$ leaves. The tree is of depth $10$ and one of
the longest pseudo-cube-free words is
$\mtt0\mtt0\mtt1\mtt1\mtt0\mtt1\mtt1\mtt0\mtt0$.

Suppose $\theta=(\mtt0,\mtt1)\cdot\mirror$. Similarly, we verified
by computer that there are only finitely many pseudo-cube-free
words. There are in total $15$ nodes, including $8$ leaves. The tree
is of depth $3$ and one of the longest pseudo-cube-free words is
$\mtt0\mtt0$. In fact, any word in this setting is a pseudo power.
\end{proof}

\begin{proposition}\label{prop:pseudodekking}
There is a pseudo-cube-free infinite word over $3$ letters with any
antimorphic involution.
\end{proposition}
\begin{proof}
There are two kinds of antimorphic involutions over $3$ letters:
$\theta$ is either the mirror image or a transposition composed with
the mirror image.

Suppose $\theta$ is the mirror image. The following morphism $d_3$
given by Dekking \cite{Dekking1979} presents an abelian-cube-free
infinite word $d_3^\omega(\mtt0)$ over $3$ letters, which is also
pseudo-cube-free:
\begin{align*}
d_3(\mtt0)=&\mtt0\mtt0\mtt1\mtt2,\\
d_3(\mtt1)=&\mtt1\mtt1\mtt2,\\
d_3(\mtt2)=&\mtt0\mtt2\mtt2.
\end{align*}
So there is a pseudo-cube-free infinite word $d_3^\omega(\mtt0)$
over $3$ letters with respect to mirror image.

Now suppose $\theta$ is a transposition composed with the mirror
image. Without loss of generality, we assume
$\theta=(\mtt0,\mtt1)\cdot\mirror$. Consider the following morphism:
\begin{align*}
t(\mtt0)=&\mtt0\mtt2\mtt1,\\
t(\mtt1)=&\mtt1\mtt2\mtt0,\\
t(\mtt2)=&\mtt2.
\end{align*}
One can verify that the word
$z=t^\omega(0)=\mtt0\mtt2\mtt1\mtt2\mtt1\mtt2\mtt0\mtt2\mtt1\mtt2\mtt0\mtt2\mtt0\mtt2\mtt1\mtt2\mtt1\mtt2\mtt0\mtt2\mtt0\mtt2\mtt1\cdots$
is the Thue-Morse sequence \cite{Thue1912} with letter $\mtt2$
inserted between every two consecutive letters. Now we prove that
$z$ is pseudo-cube-free. Suppose $z=xw_1w_2w_3y$ contains a
pseudo-cube $w_1w_2w_3$ with $\abs{w_1}=\abs{w_2}=\abs{w_3}$. Either
the last letter of $w_1$ or the first letter of $w_2$ is $\mtt2$,
but not both. Since $\theta(\mtt2)=\mtt2$, we have
$w_1\neq\theta(w_2)$. So $w_1=w_2$. By the same reason, $w_2=w_3$.
Then the length of $w_1=w_2=w_3$ must be even. Otherwise, either the
first letter of $w_1$ or the first letter of $w_2$ is $\mtt2$, but
not both, and thus we have $w_1\neq w_2$. Now since
$\abs{w_1}=\abs{w_2}=\abs{w_3}$ is even, we can omit the letter
$\mtt2$ from each word and get new words $w_1', w_2', w_3'$ such
that $w_1'=w_2'=w_3'$ and $w_1'w_2'w_3'$ is a factor of the
Thue-Morse sequence, which contradicts the fact that Thue-Morse
sequence is cube-free. Therefore, $z=t^\omega(0)$ is
pseudo-cube-free with respect to $(\mtt0,\mtt1)\cdot\mirror$ over
$3$ letters.
\end{proof}

\begin{theorem}
There is no pseudo-cube-free infinite word over $k$ letters for
$k\leq 2$, and there is a pseudo-cube-free infinite word over $k$
letters for $k\geq 3$, with respect to arbitrary antimorphic
involution.
\end{theorem}
\begin{proof}
(1) Since $2$ is the minimal size of alphabet over which there is a
cube-free infinite word, by Lemma~\ref{lemma:lemma0}, there is no
pseudo-cube-free infinite word over $k$ letters for $k\leq 1$ and
there is pseudo-cube-free infinite word with respect to $\theta$ for
$\tr(\theta)\geq 2$.

(2) There is an abelian-cube-free infinite word $d_3^\omega(\mtt0)$
over $3$ letters. By Lemma~\ref{lemma:lemma1}, there is a
pseudo-cube-free infinite word over $5$ or more letters.

(3) By Proposition~\ref{prop:pseudocube}, there is no
pseudo-cube-free infinite word over binary alphabet. By
Proposition~\ref{prop:pseudodekking}, there is a pseudo-cube-free
infinite word over $3$ letters. In particular, there is a
pseudo-cube-free infinite word over $3$ letters with respect to
mirror image, where $\abs{\id(\mirror)}=3,\abs{\tr(\mirror)}=0$. So
by Lemma~\ref{lemma:lemma2}, there is a pseudo-cube-free infinite
word with respect to $\theta$ such that
$\abs{\id(\theta)}+\abs{\tr(\theta)}\geq 3$.

The result is summarized in Table~\ref{table:pseudocube}, where the
subscription presents which situation of the (1), (2), (3) the case
falls in.
\end{proof}

\begin{table}
\centering \caption{The existence of pseudo-cube-free infinite
words}
\begin{tabular}{c|p{5.5ex}p{5.5ex}p{5.5ex}p{5.5ex}p{5.5ex}p{5.5ex}p{5.5ex}p{5.5ex}}
  \hline
  $\abs{\Sigma}$ & $1$ & $2$ & $3$ & $4$ & $5$ & $6$ & $7$ & $8$ \\
  \hline
  $\tr(\theta)=0$ & $\times_{1}$ & $\times_{3}$ & $\surd_{3}$ & $\surd_{3}$ & $\surd_{2,3}$ & $\surd_{2,3}$ & $\surd_{2,3}$ & $\surd_{2,3}$ \\
  $\tr(\theta)=1$ & $-$ & $\times_{3}$ & $\surd_{3}$ & $\surd_{3}$ & $\surd_{2,3}$ & $\surd_{2,3}$ & $\surd_{2,3}$ & $\surd_{2,3}$ \\
  $\tr(\theta)=2$ & $-$ & $-$ & $-$ & $\surd_{1}$ & $\surd_{1,2,3}$ & $\surd_{1,2,3}$ & $\surd_{1,2,3}$ & $\surd_{1,2,3}$ \\
  $\tr(\theta)=3$ & $-$ & $-$ & $-$ & $-$ & $-$ & $\surd_{1,2,3}$ & $\surd_{1,2,3}$ & $\surd_{1,2,3}$ \\
  $\tr(\theta)=4$ & $-$ & $-$ & $-$ & $-$ & $-$ & $-$ & $-$ & $\surd_{1,2,3}$ \\
  \hline
\end{tabular}\label{table:pseudocube}
\end{table}

We now discuss other pseudo-$k$th-power-free infinite words with
$k\geq 4$. Every word over a single letter is a power. So the unary
case is trivial and no X-free infinite word exists for X being
either $k$th-power, or abelian-$k$th-power, or pseudo-$k$th-power.

\begin{theorem}
For any integer $k\geq 4$, there is no pseudo-$k$th-power-free
infinite word over $l$ letters for $l\leq 1$, and there is a
pseudo-$k$th-power-free infinite word over $l$ letters for $l\geq
3$, with respect to arbitrary antimorphic involution. For binary
alphabet, the existence of pseudo-$k$th-power-free infinite words
depends on the antimorphic involution.
\end{theorem}
\begin{proof}
(1) There is no pseudo-power-free infinite words over unary
alphabet. Since there is a $k$th-power-free infinite word over
binary alphabet, by Lemma~\ref{lemma:lemma0}, there is a
pseudo-$k$th-power-free infinite word with respect to $\theta$ for
$\tr(\theta)\geq 2$.

(2) There exists abelian-$4$th-power-free infinite word over binary
alphabet, such as the following construction by Dekking
\cite{Dekking1979} $w=d_4^\omega(\mtt0)$ where
\begin{align*}
d_4(\mtt0)=&\mtt0\mtt1\mtt1, \\
d_4(\mtt1)=&\mtt0\mtt0\mtt0\mtt1.
\end{align*}
So there exists an abelian-$k$th-power-free infinite word $w$ over
binary alphabet for any integer $k\geq 4$. By
Lemma~\ref{lemma:lemma1}, there is a pseudo-$k$th-power-free
infinite word over $3$ or more letters.

(3) That infinite word $w=d_4^\omega(\mtt0)$ is also a
pseudo-$k$th-power-free infinite word for $k\geq 4$ over binary
alphabet with respect to the mirror image, where
$\abs{\id(\mirror)}=2,\abs{\tr(\mirror)}=0$. So by
Lemma~\ref{lemma:lemma2}, there is a pseudo-$k$th-power-free
infinite word for $k\geq 4$ with respect to $\theta$ such that
$\abs{\id(\theta)}+\abs{\tr(\theta)}\geq 2$. If
$\theta=(\mtt0,\mtt1)\cdot\mirror$ over binary alphabet, then any
word is a pseudo power.

The result is summarized in Table~\ref{table:pseudoforth}, where the
subscription presents which situation of the (1), (2), (3) the case
falls in.
\end{proof}

\begin{table}
\centering \caption{The existence of pseudo-$k$th-power-free
infinite words for integer $k\geq 4$}
\begin{tabular}{c|p{5.5ex}p{5.5ex}p{5.5ex}p{5.5ex}p{5.5ex}p{5.5ex}p{5.5ex}p{5.5ex}}
  \hline
  $\abs{\Sigma}$ & $1$ & $2$ & $3$ & $4$ & $5$ & $6$ & $7$ & $8$ \\
  \hline
  $\tr(\theta)=0$ & $\times_{1}$ & $\surd_{3}$ & $\surd_{2,3}$ & $\surd_{2,3}$ & $\surd_{2,3}$ & $\surd_{2,3}$ & $\surd_{2,3}$ & $\surd_{2,3}$ \\
  $\tr(\theta)=1$ & $-$ & $\times_{3}$ & $\surd_{2,3}$ & $\surd_{2,3}$ & $\surd_{2,3}$ & $\surd_{2,3}$ & $\surd_{2,3}$ & $\surd_{2,3}$ \\
  $\tr(\theta)=2$ & $-$ & $-$ & $-$ & $\surd_{1,2,3}$ & $\surd_{1,2,3}$ & $\surd_{1,2,3}$ & $\surd_{1,2,3}$ & $\surd_{1,2,3}$ \\
  $\tr(\theta)=3$ & $-$ & $-$ & $-$ & $-$ & $-$ & $\surd_{1,2,3}$ & $\surd_{1,2,3}$ & $\surd_{1,2,3}$ \\
  $\tr(\theta)=4$ & $-$ & $-$ & $-$ & $-$ & $-$ & $-$ & $-$ & $\surd_{1,2,3}$ \\
  \hline
\end{tabular}\label{table:pseudoforth}
\end{table}

\section{Testing pseudo-power-freeness of words}\label{section:decision}
In this section, we will discuss the following problem: given a
finite word $w$ and integer $k\geq 2$, does $w$ contain any
pseudo-$k$th-power as a factor? First, we will discuss the general
algorithm for arbitrary $k$.

\subsection{General algorithm for arbitrary $k$th pseudo-power}
The na\"\i ve algorithm runs in $O(N^3)$ time to decide whether $w$
contains any pseudo-$k$th-power as a factor. The idea is that we
check each possible candidate factors $u$ of $w$ to see whether $u$
is a pseudo-$k$th-power. There are $O(N^2)$ factors and check
whether a word is a pseudo $k$th power can be done with $O(N)$
comparisons of letters.

    Here we describe an $O(n^2 \lg n)$-time algorithm to decide weather an input string $w$ of length $n$
    contains a $k$-th pseudo-power of a word or not.
    Our algorithm has three steps: in the first step, it constructs an $n\times n\times\log
    n\times2$ zero-one matrix $A$ such that
    \[
      \begin{array}{l}
      A_{i, j, k, 0} = 1 \textrm{ iff } w[i\,..\, i+2^k-1] = w[j\,..\, j+2^k-1],\\
      A_{i, j, k, 1} = 1 \textrm{ iff } w[i\,..\, i+2^k-1] = \phi(w[j\,..\, j+2^k-1]).
      \end{array}
    \]
    Then, using $A$, the algorithm constructs a set of binary strings
    \[\{s_i: 1\leq i \leq \left\lfloor \frac n 2\right\rfloor, |s_i|=n-2i+1, s_i[\ell]=1 \textrm{
    iff } w[\ell\,..\, \ell+2i-1] \textrm{ is a pseudo-square}\}.\]
    Having $s_i$'s, it is easy to find a pseudo-$k$th-power, if there exists any.
    \begin{lemma}\label{lem:two-to-k}
      Given $\{s_i:1\leq i\leq \lfloor n/2\rfloor\}$ and $k$ as inputs, there is an algorithm with
      time linear to $\sum|s_i|$ that finds {\em all} pseudo-$k$th-powers in $w$.
    \end{lemma}
    %{\noindent\bf Proof.}
    \begin{proof}
      In linear time, for all $1\leq i\leq \lfloor n/2\rfloor$, we will break the string $s_i$ into $i$ strings $s_{i,1}, s_{i, 2}, \ldots,
      s_{i, i}$ such that $s_{i,j}$ consists of the characters at positions $j, j+i, j+2i, \ldots$ in
      $s_i$.
      This can be done trivially in linear-time.

      Now, observe that there is a pseudo-$k$th-power in $w$ starting at position $x$ of length
      $k\times y$ if and
      only if $s_{y, (x - 1 \mod y) + 1}$ has $k-1$ consecutive $1$s starting at position
      $\lceil x/y\rceil$.
    \hfill$\Box$

    Our method is summarized in Algorithm~\ref{alg:main}.

    \begin{algorithm}
      \label{alg:main}
      \caption{\textsc{Pseudo-Power-Freeness}$(w,k)$}
      Initial $A$ to a zero matrix\;
      \For{all $i$ and $j$ such that $1\leq i,j\leq n$} {
        \lIf{$w[i] = w[j]$} {$A_{i,j,0,0} \leftarrow 1$}\;
        \lIf{$w[i] = \phi(w[j])$} {$A_{i,j,0,1} \leftarrow 1$}\;
      }
      \SetVline
      \nl\For{$k=1\ldots \lfloor\lg n\rfloor$}{
        \SetNoline
        \For{all $i$ and $j$ such that $1\leq i, j, \leq n-2^k+1$} {
          \lIf{$A_{i,j,k-1,0} = 1$ and $A_{i+2^{k-1},j+2^{k-1},k-1,0}=1$} {$A_{i,j,k,0} \leftarrow 1$}\;
          \lIf{$A_{i,j+2^{k-1},k-1,1} = 1$ and $A_{i+2^{k-1},j,k-1,0}=1$} {$A_{i,j,k,1} \leftarrow 1$}\;
        }
      }
      \SetVline
      \nl\For{all $i$ such that $1\leq i\leq\lfloor n/2\rfloor$} {
        \SetNoline
        $s_i \leftarrow 0^{n-2i}$\;
        \For{$1\leq \ell\leq n-2i$}{
          Let $i_1, i_2, \ldots, i_I$ be distinct integers such that $i=\sum_{x=1}^I 2^{i_x}$\;
          \lIf{$A_{\ell+\sum_{x=1}^{y-1}2^{i_x}, \ell+i+\sum_{x=1}^{y-1}2^{i_x}, i_y, 0} = 1$
            for all $y=1,2,\ldots,I$}{$s_i[\ell] \leftarrow 1$\;}
          \lIf{$A_{\ell+\sum_{x=1}^{y-1}2^{i_x}, \ell+2i-\sum_{x=1}^y2^{i_x}, i_y, 1} = 1$
            for all $y=1,2,\ldots,I$}{$s_i[\ell] \leftarrow 1$\;}
        }
      }
      \SetVline
      \nl\For{all $i$ such that $1\leq i\leq\lfloor n/2\rfloor$} {
        \SetNoline
        Break $s_i$ into $i$ strings $s_{i,1}, s_{i,2}, \ldots, s_{i,i}$ as described in
        Lemma~\ref{lem:two-to-k}\;
        \lIf{there exists $1\leq j\leq i$ such that $s_{i,j}$ contains $k-1$ consecutive $1$s}
        {\Return NO\;}
      }

      \Return YES\;
    \end{algorithm}

  \begin{theorem}
    Algorithm~\ref{alg:main} runs in time $O(n^2\lg n)$ and returns YES if and only if $w$ does not have any pseudo-$k$th-power as
    a substring.
  \end{theorem}
  {\noindent\bf Proof.}
    The running time of block 1 and 2 is $O(n^2\lg n)$ and block 3 runs in $O(n^2)$ as explained in
    Lemma~\ref{lem:two-to-k}.

    As for the correctness, it is enough to show that, after block 1 and 2, the matrix $A$ and the set of strings
    $\{s_i:1\leq i\leq \lfloor n/2\rfloor\}$ have the values that the are supposed to have; i.e.
    \begin{eqnarray}
      A_{i, j, k, 0} = 1 &\textrm{ iff }& w[i\,..\, i+2^k-1] = w[j\,..\, j+2^k-1],\\
      A_{i, j, k, 1} = 1 &\textrm{ iff }& w[i\,..\, i+2^k-1] = \phi(w[j\,..\, j+2^k-1]),
    \end{eqnarray}
    and
    \begin{eqnarray}
      \forall i,\ell \textrm{ s.t. } 1\leq\ell\leq i: s_i[\ell]=1 &\textrm{ iff }& w[\ell\,..\, \ell+2i-1] \textrm{ is a pseudo-square}.
    \end{eqnarray}
    To prove that $(1)$ holds, we use induction on $k$.
    For $k=0$ $(1)$ holds because of the initialization before block 1.
    Assuming that $(1)$ holds for $k=0,1,\ldots,k'$, it is easy to see that $(1)$ holds for $k'+1$:
    $w[i\ldots i+2^{k'+1}-1] = w[j\ldots j+2^{k'+1}-1]$ if and only if $w[i\ldots i+2^{k'}-1] =
    w[j\ldots j+2^{k'}-1]$ and $w[i + 2^{k'}\ldots i+2^{k'+1}-1] = w[j+2^{k'}\ldots j+2^{k'+1}-1]$.

    Proving $(2)$ is similar to proving $(1)$.

    For $(3)$, note that
    \begin{enumerate}
      \item $w[\ell\,..\, \ell+2i-1]$ is a pseudo-square if and only if $w[\ell\,..\,\ell+i-1] =
        w[\ell+i\,..\,\ell+2i-1]$ or $w[\ell\,..\,\ell+i-1] = \phi(w[\ell+i\,..\,\ell+2i-1])$.

      \item $w[\ell\,..\,\ell+i-1] = w[\ell+i\,..\,\ell+2i-1]$ if and only if
        $w[\ell+\sum_{x=1}^{y-1}2^{i_x}\,..\,\ell+\sum_{x=1}^y2^{i_x}-1] =
        w[\ell+i+\sum_{x=1}^{y-1}2^{i_x}\,..\,\ell+i+\sum_{x=1}^y2^{i_x}-1]$ for all
        $y=1,2,\ldots,I$, where $i_1, i_2, \ldots, i_I$ are distinct integers such that $i=\sum_{x=1}^I2^{i_x}$.

      \item $w[\ell\,..\,\ell+i-1] = \phi(w[\ell+i\,..\,\ell+2i-1])$ if and only if
        $w[\ell+\sum_{x=1}^{y-1}2^{i_x}\,..\,\ell+\sum_{x=1}^y2^{i_x}-1] =
        \phi(w[\ell+2i-\sum_{x=1}^y2^{i_x}\,..\,\ell+2i-\sum_{x=1}^{y-1}2^{i_x}-1])$ for all
        $y=1,2,\ldots,I$, where $i_1, i_2, \ldots, i_I$ are distinct integers such that
        $i=\sum_{x=1}^I2^{i_x}$.\qedhere
    \end{enumerate}
  %\hfill$\Box$
  \end{proof}

In the following subsection, we consider, for fixed small $k$,
whether a given word $w$ is pseudo-$k$th-power-free.

\subsection{Testing pseudo-square-freeness}
\begin{theorem}
To decide whether a word $w$ contains a pseudo-square as a factor
can be done in linear time.
\end{theorem}
\begin{proof}
Let $N=\abs{w}$. A word $w$ contains a pseudo-square if and only if
$w$ contains a square or a word of the form $w\phi(w)$.

To check whether $w$ contains a square can be done in linear time.
There are a few works in the literatures on testing square-freeness
in linear time \cite{Crochemore1983,Main&Lorentz1985}.

To check whether $w$ contains a word of the form $u\phi(u)$, it is
enough to check whether $w$ contains a word $a\phi(a)$ for a letter
$a$. To see this, if w contains $u\phi(u)$, then let $a$ be the
right-most letter of $u$ and $w$ contains $a\phi(a)$; for the other
direction, word $a\phi(a)$ itself is a pseudo-square.

\begin{algorithm}
  \SetLine \SetKw{KwFrom}{from} \SetKw{KwBreak}{break}  \SetKw{KwContinue}{continue}
  \linesnumbered
  \KwIn{a word $w=w[1\,..\,N]$.}
  \KwOut{``YES'' if $w$ is pseudo-square-free; ``NO'' otherwise.}
  \lIf{$w$ contains a square}{\Return{``NO''}}\;
  \For{$i$ \KwFrom $1$ \KwTo $N-1$}{
    \lIf{$\phi(w[i])=w[i+1]$}{\Return{``NO''}}\;
  }
  \Return{``YES''}\;
  \caption{Decide whether $w$ is pseudo-square-free in linear time}
  \label{figure:square}
\end{algorithm}

The algorithm is illustrated in Algorithm~\ref{figure:square}. It is
obvious that the algorithm is linear.
\end{proof}

\subsection{Testing pseudo-cube-freeness}
Before we show a cubic time algorithm for the pseudo-cube-freeness
of a word, we first introduce some concepts. Let $w=w[1\,..\,N]$ be
a finite word over $\Sigma$ and let $\phi$ be an antimorphic
involution with the same alphabet $\Sigma$. A \emph{right minimal
periodic} $rmp_w[1\,..\,N]$ of $w$ is a vector and is defined by
  \[rmp_w[i]=\min\left\{\min\{n:w[i\,..\,i+2n-1]=x^2\textrm{ for some }x\neq\epsilon\},+\infty\right\},\]
and similarly a \emph{left minimal periodic} $lmp_w[1\,..\,N]$ of
$w$ is defined by
  \[lmp_w[i]=\min\left\{\min\{n:w[i-2n+1\,..\,i]=x^2\textrm{ for some }x\neq\epsilon\},+\infty\right\}.\]
For example, when $w=\tt01001010$, we have
$rmp_w=[3,+\infty,1,2,2,+\infty,+\infty,+\infty]$ and
$lmp_w=[+\infty,+\infty,+\infty,1,+\infty,3,2,+\infty]$. A
\emph{centralized maximal pseudo-palindrome} $cmp_w[0\,..\,N]$ of
$w$ (with respect to $\phi$) is a vector and is defined by
  \[cmp_w[i]=\max\left\{\max\{m:\phi(w[i-m\,..\,i-1])=w[i\,..\,i+m-1]\},0\right\}.\]
For example, when $w=\tt01001010$ and $\phi=\mirror$, we have
$cmp_w=[0,0,0,3,0,0,0,0,0]$. The left-most and right-most elements
of $cmp_w$ are always $0$.

\begin{lemma}\label{lemma:rmplmp}
For any fixed integer $k$, the right (respectively, left) minimal
periodic $rmp_w$ (respectively, $lmp_w$) of word $w$ can be computed
in linear time $O(\abs{w})$.
\end{lemma}
There is an algorithm to compute $rmp_w$, the shortest square
starting at each position, in linear time \cite{Kosaraju1994} by
using suffix tree. Since vector $lmp_w$ can be obtained by first
computing $V=rmp_{\mirror(w)}$ and then reversing $V$, vector
$lmp_w$ can also be computed in linear time.

\begin{lemma}\label{lemma:cmp}
The centralized maximal pseudo-palindrome $cmp_w$ can be computed in
linear time $O(\abs{w})$.
\end{lemma}
Lemma~\ref{lemma:cmp} has been proved in the book
\cite[page~197--198]{Gusfield1997}, which claimed all the maximal
even-length palindromes can be found in linear time.

Now we are ready to show a quadratic time algorithm to test the
pseudo-cube-freeness of a given word $w$ of length $N$. By
definition, a pseudo-cube is in one of the following form $xxx$,
$xx\phi(x)$, $\phi(x)xx$, and $x\phi(x)x$. In order to check whether
$w$ contains any pseudo-cube, we check each of the four cases.

To check whether $w$ contains any word of the form $xxx$ can be done
in linear time. Word $w$ contains a cube if and only if one of the
maximal repetition in $w$ has exponent $\geq 3$, and there is linear
algorithm \cite{Kolpakov&Kucherov1999} to find all the maximal
repetitions. So this case can be checked in $O(N)$ time.

To check whether $w$ contains any word of the form $xx\phi(x)$, we
check whether there is a pair of factors $w[i-2n+1\,..\,i]=yy$ and
$w[i-m+1\,..\,i+m]=z\phi(z)$ that overlap in the sense that $n\leq
m$. By the definitions of $lmp_w$ and $cmp_w$, we only need to check
for each position $i$ whether $lmp_w[i]\leq cmp_w[i]$. this can be
done in $O(N)$ time when all $lmp_w,cmp_w$ are already computed. The
case for $\phi(x)xx$ is similar.

To check whether $w$ contains any word of the form $x\phi(x)x$, we
check whether there is a pair of factors
$w[i-n+1\,..\,i+n]=y\phi(y)$ and $w[j-m+1\,..\,j+m]=z\phi(z)$ that
overlap in the sense that $\abs{i-j}\leq n$ and $\abs{i-j}\leq m$.
By the definition of $cmp_w$, we check for each pair of indices
$i,j$ with $i<j$ whether $j-i\leq cmp_w[i]$ and $j-i\leq cmp_w[j]$.
This can be done in $O(N^2)$ time when $cmp_w$ is already known.

\begin{algorithm}
  \SetLine \SetKw{KwFrom}{from}
  \linesnumbered
  \KwIn{a word $w=w[1\,..\,N]$.}
  \KwOut{``YES'' if $w$ is pseudo-cube-free; ``NO'' otherwise.}
  compute $rmp_w$, $lmp_w$, $cmp_w$\;
  \lIf(\tcp{The case $xxx$}){$w$ contains a cube}{\Return{``NO''}}
  \For{$i$ \KwFrom $1$ \KwTo $N$}{
    \lIf(\tcp{The case $\phi(x)xx$}){$rmp_w[i]\leq cmp_w[i-1]$}{\Return{``NO''}}
    \lIf(\tcp{The case $xx\phi(x)$}){$lmp_w[i]\leq cmp_w[i]$}{\Return{``NO''}}
    \For{$d$ \KwFrom $1$ \KwTo $cmp_w[i]$}{
      \lIf(\tcp{The case $x\phi(x)x$}){$d\leq cmp_w[i+d]$}{\Return{``NO''}}}
  }
  \Return{``YES''}\;
  \caption{Decide whether $w$ is pseudo-cube-free in linear time}
  \label{figure:cube}
\end{algorithm}
The completed algorithm is given in Algorithm~\ref{figure:cube}. So
we have the following theorem.
\begin{theorem}
To decide whether a word $w$ contains a pseudo-cube as a factor can
be done in quadratic time.
\end{theorem}
\begin{proof}
Let $N=\abs{w}$. Algorithm~\ref{figure:cube} checks the
pseudo-cube-freeness of $w$ in quadratic time. By
Lemma~\ref{lemma:rmplmp} and Lemma~\ref{lemma:cmp}, the computation
of $rmp_w, lmp_w, cmp_w$ in line~1 can be done in $O(N)$ times.
Line~2 can be done in $O(N)$ time. Line~3--10 can be done in
$O(N^2)$ time. So the algorithm runs in $O(N^2)$ time.

Now we prove the correctness of the algorithm. First, we prove that
if the algorithm returns ``NO'', then $w$ contains a pseudo cube. If
the algorithm return at line~2, then $w$ contains a cube of the form
$xxx$, which is also a pseudo cube. Suppose the algorithm return at
line~4. Let $n=rmp_w[i]$ and $m=cmp_w[i-1]$. Then $n\leq m$ and the
word $w[i-n\,..\,i+2n-1]$ is of the form $\phi(x)xx$, which is a
pseudo cube. Suppose the algorithm return at line~5. Let
$n=lmp_w[i]$ and $m=cmp_w[i]$. Then $n\leq m$ and the word
$w[i-2n+1\,..\,i+n]$ is of the form $xx\phi(x)$, which is a pseudo
cube. Suppose the algorithm return at line~7. Then the word
$w[i-d+1\,..\,i+2d]$ is of the form $x\phi(x)x$, which is a pseudo
cube.

Now, we prove that if $w$ contains a pseudo cube, then the algorithm
returns ``NO''. If $w$ contains a pseudo cube of the form $xxx$,
then the algorithm returns at line~2. Suppose
$w[s\,..\,s+3p-1]=xx\phi(x)$. Then $q=lmp_w[s+2p-1]\leq\abs{x}$ and
% $\mirror(x)$ is a suffix of (\mirror(w[s+2p-q\,..\,s+2p-1]))^\infty$. In addition,
$cmp_w[s+2p-1]\geq\abs{x}\geq q$. So the algorithm returns at line~5
for $i=s+2p-1$, (although the detected pseudo cube
$w[s+2p-2q\,..\,s+2p+q-1]$ may be different from
$w[s\,..\,s+3p-1]$.) The case $w[s\,..\,s+3p-1]=\phi(x)xx$ is
similar. Suppose $w[s\,..\,s+3p-1]=x\phi(x)x$. Then
$cmp_w[s+p-1]\geq\abs{x}$ and $cmp_w[s+2p-1]\geq\abs{x}$. So the
algorithm returns at line~7 for $i=s+p-1$ and $d=p$.
\end{proof}

\section{Conclusion}\label{section:conclusion}
In this paper, we discussed the existence of infinite words that do
not contain pseudo-$k$th-power. For alphabet size $\leq 2$, there is
no pseudo-square-free infinite words and for alphabet size $\geq 5$,
there exist pseudo-square-free infinite words. For other alphabet
size, the existence of pseudo-square-free infinite words depends on
the antimorphic involution $\phi$. For alphabet size $\leq 2$, there
is no pseudo-cube-free infinite words and for alphabet size $\geq
3$, there exist pseudo-cube-free infinite words. For alphabet size
$\geq 3$, there exist pseudo-$k$th-power-free infinite words for any
integer $k\geq 4$. For binary alphabet and any integer $k\geq 4$,
the existence of pseudo-$k$th-power-free infinite words depends on
the antimorphic involution $\phi$.

We also discussed the algorithm for testing whether a given word $w$
is pseudo-$k$th-power-free. For arbitrary $k$th pseudo-power, there
is a $O(n^2\lg n)$-time algorithm to find all $k$th pseudo-power in
$w$, where $n=\abs{w}$. For $k=2$, there is a $O(n)$-time algorithm
for testing pseudo-square-freeness of word $w$ of length $n$. For
$k=3$, there is a $O(n^2)$-time algorithm for testing
pseudo-cube-freeness of word $w$ of length $n$.

\section*{Acknowledgement}
The authors wish to thank Dr. Shinnosuke Seki and Professor Lucian
Ilie for very helpful discussion and comments on the draft of this
paper.

% \bibliographystyle{plain}
% \bibliography{xppowers}

\end{document}